# Multifractal analysis of charged particle multiplicity distribution in the framework of Renyi entropy

Swarnapratim Bhattacharyya[1], Maria Haiduc[2], Alina Tania Neagu[2] and Elena Firu[2]

[1]Department of Physics, New Alipore College, L Block, New Alipore, Kolkata 700053, India

Email: swarna_pratim@yahoo.com

[2]Institute of Space Science, Bucharest, Romania

## Abstract

A study of multifractality and multifractal specific heat has been carried out for the produced shower particles in nuclear emulsion detector for $^{16}$O-AgBr, $^{28}$Si-AgBr and $^{32}$S-AgBr interactions at 4.5AGeV/c in the framework of Renyi entropy. Experimental results have been compared with the prediction of Ultra Relativistic Quantum Molecular Dynamics (UrQMD) model. Our analysis reveals the presence of multifractality in the multiparticle production process in high energy nucleus-nucleus interactions. Degree of multifractality is found to be higher for the experimental data and it increases with the increase of projectile mass. The investigation of quark-hadron phase transition in the multiparticle production in $^{16}$O-AgBr, $^{28}$Si-AgBr and $^{32}$S-AgBr interactions at 4.5 AGeV/c in the framework of Ginzburg-Landau theory from the concept of multifractality has also been presented. Evidence of constant multifractal specific heat has been obtained for both experimental and UrQMD simulated data.

# 1. Introduction

The study of non statistical fluctuations and correlations in relativistic and ultra-relativistic nucleus-nucleus collisions have become a subject of major interest among the particle physicists. Bialas and Peschanski [1-2] proposed a new phenomenon called intermittency to study the non-statistical fluctuations in terms of the scaled factorial moment. In high-energy physics intermittency is defined as the power law behavior of scaled factorial moment with the size of the considered phase space [1-2]. This method has its own advantage that it can extract non-statistical fluctuations after extricating the normal statistical noise [1-2]. The study of non-statistical fluctuations by the method of scaled factorial moment leads to the presence of self-similar fractal structure in the multi-particle production of high energy nucleus-nucleus collisions [3]. The self-similarity observed in the power law dependence of scaled factorial moments reveals a connection between intermittency and fractality. The observed fractal structure is a consequence of self-similar cascade mechanism in multiparticle production process. The close connection between intermittency and fractality prompted the scientists to study fractal nature of multiparticle production in high energy nucleus- nucleus interactions. To get both qualitative and quantitative idea concerning the multiparticle production mechanism fractality study in heavy ion collisions is expected to be very resourceful.

# 2. Fractals-Multifractals and Monofractals

The term "fractal" was coined by Mandelbrot [4] from the Latin word fractus which means broken or fractured. Mandelbrot introduced the new geometry called the fractal geometry to look into the world of apparent irregularities. A fractal pattern is one that scales infinitely to reproduce itself such that the traditional geometry does not define it. In other words a fractal is generally a rough or fragmented geometrical shape that can be split into parts, each of which is at least approximately reduced size copy of the whole [4]. Sierpinski triangle, Koch snow flack, Peano curve, Mandelbrot set , Lorentz attractor are the well known mathematical structures that exhibit fractal geometry [4]. Fractals also describe many real world objects such as clouds, mountains, coastlines etc those do not corresponds to simple geometrical shape [4]. Fractals can be classified into two categories: multifractals and monofractals [4]. Multifractals are complicated self-similar objects that consist of differently weighted fractals with different non-integer dimensions. The fundamental characteristic of multifractality is that the scaling properties may be different in different regions of the systems [4].

Monofractals are those whose scaling properties are the same in different regions of the system [4]. As the scaling properties are dissimilar in different parts of the system, multifractal systems require at least more than one scaling exponent to describe the scaling behavior of the system [5]. A distinguishing feature of the processes that have multifractal characteristics is that various associated probability distributions display power law properties [6-7]. Multifractal theory is essentially rooted in probability theory, though draws on complex ideas from each of physics, mathematics, probability theory and statistics [6]. Apart from multiparticle production in high energy physics, multifractal analysis has proved to be a valuable method of capturing the underlying scaling structure present in many types of systems including diffusion limited aggregation [8-10], fluid flow through random porous media [11], atomic spectra of rare-earth elements [12], cluster-cluster aggregation [13] and turbulent flow [14]. In physiology, multifractal structures have been found in heart rate variability [15] and brain dynamics [16]. Multifractal analysis has been helpful in distinguishing between healthy and pathological patients [17]. Multifractal measures have also been found in man-made phenomena such as the Internet [18], art [19] and the stock market [20-22]. Multifractals has also been used in a wide range of application areas like the description of dynamical systems, rainfall modelling, spatial distribution of earthquakes and insect populations, financial time series modelling and internet traffic modeling [6-7].

It should be mentioned here that the most important property of fractals are their dimensions [4,6-7]. Fractal dimension is used to describe the size of the fractal sets [4,6-7]. For example, the dimension of an irregular coastline may be greater than one but less than two, indicating that it is not like a simple line and has space filling characteristics in the plane. Likewise, the surface area of a snowflake may be greater than two but less than three, indicating that its surface is more complex than regular geometrical shapes, and is partially volume filling [4,6-7]. Fractal dimension can be calculated by taking the limit of the quotient of the log change in object size and the log change in measurement scale, as the measurement scale approaches to zero [4]. The differences came in what is meant by the measurement scale and how to get an average number out of many different parts of the geometrical objects [4,6-7]. Fractal dimension quantifies the static geometry of an object [4].

Generalized fractal dimension $D_q$ is a well known parameter which reflects the nature of fractal structure [4]. From the dependence of generalized fractal dimension $D_q$ on the order q a distinguishable characterization of fractality is possible [4]. Decrease of generalized fractal dimension $D_q$ with the order of moment q signals the presence of multifractality. On the other hand if $D_q$ remains constant with the increase of order q monofractality occurs. It has been pointed out in [23] that if Quark-Gluon-Plasma (QGP) state is created in hadronic collisions, a phase transition to hadronic matter will take place. The hadronic system will show monofractality in contrast to an order dependence of generalized fractal dimension $D_q$, when a cascade process occurs [23].

Hwa [24] was the first to provide the idea of using multifractal moments $G_q$, to study the multifractality and self-similarity in multiparticle production. According to the method enunciated by Hwa [24] if the particle production process exhibits self-similar behavior, the $G_q$ moments show a remarkable power law dependence on phase space bin size. However, if the multiplicity is low, the $G_q$ moments are dominated by statistical fluctuations. In order to suppress the statistical contribution, a modified form of $G_q$ moments in terms of the step function was suggested by Hwa and Pan [25]. Takagi [26] also proposed a new method called Takagi moment method ($T_q$ moment) for studying the fractal structure of multiparticle production. Both the $G_q$ and the $T_q$ techniques have been applied extensively to analyze several high-energy nucleus-nucleus collision data [27-35]. Very recently some sophisticated methods have also been applied to study the fractal nature of multiparticle production process [36-47].

### 3. Entropy and Fractality

In high-energy nucleus-nucleus collisions, entropy measurement of produced shower particles may provide important information in studying the multiparticle production mechanism [48]. In high-energy collisions, entropy is an important parameter and it is regarded as the most significant characteristic of a system having many degrees of freedom [48]. As entropy reflects how the effective degrees of freedom changes from hadronic matter at low temperature to the quark-gluon plasma state at high temperature, it is regarded as a useful probe to study the nature of deconfinement phase transition [48]. Entropy plays a key role in the evolution of the high temperature quark-gluon plasma in ultra relativistic nucleus-nucleus interactions in Relativistic Heavy Ion Collider (RHIC) experiments and in Large Hadron Collider (LHC) experiments [48]. In nucleus-nucleus collisions, entropy measurement can be used not only to search for the formation of Quark-Gluon-Plasma (QGP) state but it may also serve as an additional tool to investigate the correlations and event-by-event fluctuations [49].

Different workers have investigated the evolution of entropy in high energy nucleus-nucleus collisions at different times [**50-66**]. The entropy of produced particles can easily be calculated from the multiplicity distribution of the data. If $P_n$ is the probability of producing n particles in a high energy interaction the entropy S is defined by the relation $S = -\sum_n P_n \ln P_n$ [**67**]. Mathematically this entropy S is called the Shannon entropy [**68**].

Apart from studying the well known Shannon entropy, people are interested to explore the hidden physics of Renyi entropy [**68-71**] as well, mainly motivated by the inspiration of A. Białas and W. Czyz˙ [**66,72-73**]. According to C.W Ma and Y.G. Ma [**74**] the difference between qth order Renyi entropy and 1st order Renyi entropy is found just to be a q dependent constant but which is very sensitive to the form of probability distribution. Renyi entropy can play a potential role to investigate the fractal characteristics of multiparticle production process [**75-76**]. The advantage of this method to study the fractal properties of multiparticle production process is that it is not related to the width and resolution of the phase space interval [**75-76**]. This method can be applied to events having higher as well as lower multiplicity. This method never suffers from the drawback of lower statistics.

In terms of the probability of multiplicity distribution $\mathbf{P_n}$, the q th order Renyi entropy can be defined as [**66,72-73**]

$$\mathbf{H_q} = \frac{1}{(q-1)} \ln[\sum_n (\mathbf{P_n})^q] \quad \ldots\ldots\ldots\ldots\ldots (1)$$

If $\mathbf{C_q} = \sum_n (\mathbf{P_n})^q$, then equation (1) can be written as

$$\mathbf{H_q} = \frac{1}{(q-1)} \ln[\mathbf{C_q}] \quad \ldots\ldots\ldots (2)$$

Generalized fractal dimension $\mathbf{D_q}$ can be evaluated from the concept of Renyi entropy according to the relation $\mathbf{D_q} = \frac{\mathbf{H_q}}{\mathbf{Y_m}} \quad \ldots\ldots\ldots (3)$

Where $\mathbf{Y_m}$ is the central rapidity value in the centre of mass frame and is given by

$$\mathbf{Y_m} = \ln[(\sqrt{s} - 2\mathbf{m_n}\langle\mathbf{n_p}\rangle/\mathbf{m_\pi})] \ldots\ldots\ldots\ldots (4).$$

Here $\sqrt{s}$ is the centre of mass energy of the concerned collision process, $\mathbf{m_\pi}$ is the rest mass of pions, $\langle\mathbf{n_p}\rangle$ denotes the average number of participating nucleons.

The generalized fractal dimension $\mathbf{D_q}$ is related with the anomalous fractal dimension $\mathbf{d_q}$ by a simple mathematical relation [**77**] $\mathbf{d_q} = 1 - \mathbf{D_q} \ldots\ldots\ldots\ldots\ldots (5).$

Goal of our present study is to carry out an investigation of multifractality and multifractal specific heat in shower particle multiplicity distribution from the concept of Renyi entropy measurements in $^{16}$O-AgBr, $^{28}$Si-AgBr and $^{32}$S-AgBr interactions at 4.5AGeV/c. We have compared our experimental results with the prediction of Ultra Relativistic Quantum Molecular Dynamics (UrQMD) model. Importance of this study is that so far very few attempts have been made to explore the presence of multifractality in multiparticle production process in the framework of higher order Renyi entropy in high energy nucleus-nucleus interactions.

## 4. Experimental Details

In order to collect the data used for the present analysis, stacks of NIKFI-BR2 emulsion pellicles of dimension 20cm×10 cm×600**μm** were irradiated by the $^{16}$O, $^{28}$Si and $^{32}$S beam at 4.5 AGeV/c obtained from the Synchrophasatron at the Joint Institute of Nuclear Research (JINR), Dubna, Russia [78-81]. According to the Powell [82], in nuclear emulsion detector particles emitted and produced from an interaction are classified into four categories, namely the shower particles, the grey particles, the black particles and the projectile fragments. Details of scanning and measurement procedure of our study along with the characteristics of these emitted and produced particles in nuclear emulsion can be found from our earlier publications [78-81].

One of the problem encountered in interpreting results from nuclear emulsion is the non-homogeneous composition of emulsion which contain both light (H,C,N,O) and heavy target nuclei (Ag, Br). In emulsion experiments it is very difficult to identify the exact target nucleus [78]. Based on the number of heavy tracks ($N_h$) total number of inelastic interactions can be divided into three broad target groups H, CNO and AgBr in nuclear emulsion [78]. Detailed method of target identification has been described in our earlier publication [78]. For the present analysis we have not considered the events which are found to occur due to collisions of the projectile beam with H and CNO target present in nuclear emulsion. Our analysis has been carried out for the interactions of three different projectile $^{16}$O, $^{28}$Si and $^{32}$S at 4.5 AGeV/c with the AgBr target only. Applying the criteria of selecting AgBr events ($N_h$ >8) we have chosen 1057 events of $^{16}$O-AgBr interactions, 514 events of $^{28}$Si-AgBr and 434 events of $^{32}$S-AgBr interactions at 4.5 AGeV/c [78]. Our analysis has been performed on the shower tracks only. We have calculated the average multiplicity of shower tracks in each interaction and presented the values in table 1 [78-81].

## 5. Analysis and Results

In a very recent paper [**83**] we have investigated the Renyi entropy of second order of shower particles using $^{16}$O, $^{28}$Si and $^{32}$S projectiles on interaction with AgBr and CNO target present in nuclear emulsion at an incident momentum of 4.5 AGeV/c. In this paper we have extended our analysis of Renyi entropy to the study of fractality in multiparticle production of $^{16}$O-AgBr, $^{28}$Si-AgBr and $^{32}$S-AgBr interactions at 4.5AGeV/c.

In order to study the fractal nature of multiparticle production from the concept of Renyi entropy we have calculated the Renyi entropy values of order q=2-5 from the relation (1) and (2) for all the three interactions. The calculated values of Renyi entropy of different orders of the produced shower particles in $^{16}$O-AgBr, $^{28}$Si-AgBr and $^{32}$S-AgBr interactions at 4.5AGeV/c have been presented in table 2. It may be mentioned here that the values of the second order Renyi entropy have been taken from our recent publication [**83**]. From the table it can be noted that for all the interactions Renyi entropy values are found to decrease as the order number increases. The variation of Renyi entropy $\mathbf{H_q}$ with order q has been presented in figure1 for $^{16}$O-AgBr, $^{28}$Si-AgBr and $^{32}$S-AgBr interactions. Error bars drawn to every experimental point are statistical errors only. Using equation (3) and (4) we have calculated the values of generalized fractal dimension $\mathbf{D_q}$ for $^{16}$O-AgBr, $^{28}$Si-AgBr and $^{32}$S-AgBr interactions. Calculated values of generalized fractal dimension $\mathbf{D_q}$ for $^{16}$O-AgBr, $^{28}$Si-AgBr and $^{32}$S-AgBr interactions have been presented in table 3. From the table it may be noted that the values of generalized fractal dimension $\mathbf{D_q}$ decreases with the increase of order number suggesting the presence of multifractality in multipion production mechanism. Presence of multifractality during the production of shower particles indicates the occurrence of cascade mechanism in particle production process. From the table 3 it may also be noted that the values of the generalized fractal dimension remain almost the same within the experimental error for $^{16}$O-AgBr, and $^{28}$Si-AgBr interactions. But for $^{32}$S-AgBr interactions the $\mathbf{D_q}$ values are higher in comparison to the other two interactions. The variation of $\mathbf{D_q}$ against the order q has been shown in figure 2 for $^{16}$O-AgBr, $^{28}$Si-AgBr and $^{32}$S-AgBr interactions.

Now it will be interesting to see how the analysis will look like if one uses Shannon entropy instead of Renyi entropy. However, Shannon entropy cannot be calculated for different orders and hence we can only calculate the values of information fractal dimension for $^{16}$O-AgBr, $^{28}$Si-AgBr and $^{32}$S-AgBr interactions. We have calculated the values of Shannon entropies derived from the concept of Gibbs-Boltzmann theories of entropy and tabulated the values in table 4. The values of information fractal dimension for the three interactions have also been calculated and presented in the same table. Comparing table 2 and table 4 it may be noticed that the values of Shannon entropy for $^{16}$O-AgBr, $^{28}$Si-AgBr and $^{32}$S-AgBr interactions are little higher than those of Renyi entropies.

From the values of the generalized fractal dimension calculated from Renyi entropy values we have evaluated the values of anomalous fractal dimension and hence the ratio of $\frac{d_q}{d_2}$ has been calculated. In table 5 the calculated values of $\frac{d_q}{d_2}$ have been presented for $^{16}$O-AgBr, $^{28}$Si-AgBr and $^{32}$S-AgBr interactions. It is worthwhile to point out that the spiky structure of density distribution of shower particles can also be investigated with the help of a set of bunching parameters [**93**]. The higher order bunching parameters can be expressed in terms of lower order parameters resulting a linear expression for the anomalous fractal dimension [**84**].

$$\mathbf{d_q = (1-r)d_2 + \frac{q}{2}d_2 r} \quad\dots\dots\dots\dots\dots\dots\dots\dots(6)$$

So that $\frac{d_q}{d_2} = \mathbf{(1-r) + \frac{qr}{2}}$ ..............................(7)

A non zero value of the **r** implies the multifractal behaviour [**84**]. We have applied this theory to our study in order to quantify the fractal nature of shower particle production. We have plotted the variation of $\frac{d_q}{d_2}$ against the order number **q** in figure 3 for $^{16}$O-AgBr, $^{28}$Si-AgBr and $^{32}$S-AgBr interactions. The calculated values of $\mathbf{r\ from}$ the slope parameter have been presented in table 6 for our data. The value of **r** signifies the degree of multifractality. From the table it may be seen that for all the interactions the **r** value is greater than zero. This reconfirms the multifractal nature of multiparticle production mechanism. Moreover, **r** value characterizing the degree of multifractality is found to depend on the mass number of the projectile beam. Degree of multifractality is found to increase with the increase of projectile mass as evident from table 6.

R.C Hwa suggested that [**85**] from the concept of multifractality a qualitative and quantitative investigation of quark-hadron phase transition in high energy nucleus-nucleus collisions is possible. In analogy with the photo count problem at the onset of lasing in non linear optics, the coherent state description in high energy nucleus-nucleus interactions can be used in the frame work of Ginzburg-Landau theory [**85-86**]. A quantity $\mathbf{\beta_q}$ in terms of the ratio of higher order anomalous fractal dimension to the second order anomalous fractal dimension can be defined by the following relation [**85-86**]

$$\boldsymbol{\beta_q} = \frac{d_q}{d_2}(q-1) \quad\quad\quad\quad\quad (8)$$

According to Ginzburg-Landau model [**86**] $\boldsymbol{\beta_q}$ is related with (q-1) by the relation

$$\boldsymbol{\beta_q} = (q-1)^v \quad\quad\quad\quad\quad (9).$$

The relation (8) and relation (9) are found to be valid for all systems which can be described by the Ginzburg-Landau (GL) theory and also is independent of the underlying dimension of the parameters of the model [**86**]. If the value of the scaling exponent $v$ is equal to or close enough (within the experimental error) to 1.304 then a quark-hadron phase transition is expected for the experimental data [**86**]. If the measured value of $v$ is different from the critical value 1.304 considering the experimental errors then the possibility of the quark-hadron phase transition has to be ruled out [**86**]. $v$ is a universal quantity valid for all systems describable by the Ginzburg-Landau (GL) theory. It is independent of the underlying dimension or the parameters of the model. The critical exponent $v$ is an important parameter to investigate the possibility of quark-hadron phase transition, since neither the transition temperature nor the other important parameters are known there [**3, 86**]. If the signature of quark-hadron phase transition depends on the details of the heavy-ion collisions e.g. nuclear sizes, collision energy, transverse energy etc, even after the system has passed the thresholds for the creation of quark-gluon plasma, such a signature is likely to be sensitive to the theoretical model used [**3, 86**]. But the critical exponent v is independent of such details. The value of the critical exponent depends only on the validity of the Ginzburg-Landau (GL) description of the phase transition for the problem concerned. Here lies the importance of this critical exponent.

We have calculated the values of $\boldsymbol{\beta_q}$ using relation (8) and presented the values in table 5 for $^{16}$O-AgBr, $^{28}$Si-AgBr and $^{32}$S-AgBr interactions. In order to search for the Ginzburg-Landau second order phase transition we have studied the variation of $\boldsymbol{\beta_q}$ with (q-1) in figure 4, in figure 5 and in figure 6 in case of $^{16}$O-AgBr, $^{28}$Si-AgBr and $^{32}$S-AgBr interactions at 4.5 AGeV/c for the experimental data. The variations of $\boldsymbol{\beta_q}$ with (q-1) have been fitted with the function $\boldsymbol{\beta_q} = (q-1)^v$ in order to extract the critical exponent $v$. In table 7 we have shown the calculated values of the critical exponent $v$ for $^{16}$O-AgBr, $^{28}$Si-AgBr and $^{32}$S-AgBr interactions for experimental events. Table 7 reflects that the calculated values of critical exponents obtained from our analysis increase with the increase of projectile mass. The experimental values of the critical exponent $\nu$ for $^{16}$O-AgBr and $^{28}$Si-AgBr interactions are found to be lower than the critical value 1.304 while for $^{32}$S-AgBr interaction the critical exponent is higher than the critical value signifying the absence of quark-hadron phase transition in our data. Interpretation of multifractality from the thermodynamical point of view allows us to study the fractal properties of stochastic processes with the help of standard concept of thermodynamics. In thermodynamics the constant specific heat approximation is widely applicable in many important cases; for example, the specific heat of gases and solids is constant, independent of temperature over a larger or smaller temperature interval [87]. This approximation is also applicable to multifractal data of multiparticle production processes proposed by Bershadskii [88]. Bershadskii pointed out that [89] regions of the temperature where the constant-specific-heat approximation is applicable are usually far away from the phase-transition regimes. For the considered interactions the phase transition like phenomena can occur in the vicinity of q=0, where q is the inverse of temperature. Such situation is expected at the onset of chaos of the dynamical attractors [89]. Barshadskii argued that [88] multiparticle production in high energy nucleus-nucleus collisions is related to phase-transition-like phenomena [3]. He introduced a multifractal Bernoulli distribution which appears in a natural way at the morphological phase transition from monofractality to multifractality. Multifractal Bernoulii distribution plays an important role in multiparticle production at higher energies. It has been pointed out that [90] multifractal specific heat can be derived from the relation $\mathbf{D_q} = \mathbf{D_\infty} + \frac{C \ln q}{(q-1)}$ if the monofractal to multifractal transition is governed by the Bernoulli distribution. The slope of the linear best fit curve showing the variation of $\mathbf{D_q}$ against $\frac{\ln q}{(q-1)}$ has been designated as multifractal specific heat. The gap at the multifractal specific heat at the multifractality to monofractality transition allows us to consider this transition as a thermodynamic phase transition [91-92].

We have studied the variation of $\mathbf{D_q}$ against $\frac{\ln q}{(q-1)}$ for $^{16}$O-AgBr, $^{28}$Si-AgBr and $^{32}$S-AgBr interactions in figure 7. The experimental points have been fitted with the best linear behavior and the slope of the best linear behavior reflects the values of multifractal specific heat. The linear behavior in figure 7 indicates approximately good agreement between the experimental data and the multifractal Bernoulii representation. The calculated values of the multifractal specific heat for all the three interactions have been presented in table 8. From table 8 it may be noted that within the experimental error multifractal specific heat remains almost constant for the three interactions.

The experimental results have been compared with those obtained by analyzing events generated by the Ultra Relativistic Quantum Molecular Dynamics (UrQMD) model. UrQMD is a hadronic transport model and this model can be used in the entire available range of energies to simulate nucleus- nucleus interactions. For more details about this model, readers are requested to consult [**93, 94**]. In our earlier papers [**81, 83**] we have utilized the UrQMD model to simulate $^{16}$O-AgBr, $^{28}$Si-AgBr and $^{32}$S-AgBr interactions at 4.5 AGeV/c. As described in our previous papers [**81, 83**] we have generated a large sample of events using the UrQMD code (UrQMD 3.3p1) for $^{16}$O-AgBr, $^{28}$Si-AgBr and $^{32}$S-AgBr interactions at 4.5AGeV/c. We have also calculated the average multiplicities of the shower tracks for all the three interactions in case of the UrQMD data sample [**81**]. Average multiplicities of the shower tracks in case of UrQMD data sample have been presented in table 1 along with the average multiplicity values of shower particles in the case of the experimental events. Table 1 shows that the average multiplicities of the shower tracks for the UrQMD events are comparable with those of the experimental values for all the interactions [**81**]. We have calculated the values of Renyi entropy of order q=2-5 for all the three interactions using the UrQMD simulated data. The calculated values of Renyi entropy has been presented in table 2 along with the experimental values. For the UrQMD simulated data also second order Renyi entropy values have been taken from our recent publication [**83**]. From the table it may be seen that the experimentally calculated values of Renyi entropy are higher than those of their UrQMD counterparts signifies the presence of more disorderness for the experimental data. The variation of $\mathbf{H_q}$ with order q for the UrQMD simulated data have been presented in figure 1 along with the experimental plots for each interaction. For the UrQMD simulated events we have calculated the values of the generalized fractal dimension $\mathbf{D_q}$ from the values of Renyi entropy using relation (3) and (4). The calculated values of generalized fractal dimension for the UrQMD simulated events of $^{16}$O-AgBr, $^{28}$Si-AgBr and $^{32}$S-AgBr interactions at 4.5AGeV/c have been presented in table 3 along with the corresponding experimental values. From the table it is seen that for the UrQMD simulated events also the generalized fractal dimension $\mathbf{D_q}$ decrease with order q signifying the presence of multifractality for the simulated events. But the values of $\mathbf{D_q}$ for the simulated events are lower than the corresponding experimental counterparts for all the interactions. The variation of $D_q$ with order q has been shown in figure 2 for $^{16}$O-AgBr, $^{28}$Si-AgBr and $^{32}$S-AgBr interactions along with the experimental plots. We have calculated the values of Shannon entropies and information dimension for the UrQMD data set of $^{16}$O-AgBr, $^{28}$Si-AgBr and $^{32}$S-AgBr interactions. The calculated values of Shannon entropy and information dimension have been presented in table 4. From table 4 it may be noticed that the Shannon entropy and information dimension for the UrQMD simulated events in case of $^{16}$O-AgBr, $^{28}$Si-AgBr and $^{32}$S-AgBr interactions are lower than the corresponding experimental values.

As in the case of experimental data in case of UrQMD simulated data also we have calculated the values of $\frac{d_q}{d_2}$ and presented the values in table 5. From the table it may be noticed that the experimentally obtained values of $\frac{d_q}{d_2}$ are little higher than the corresponding UrQMD simulated values. To quantify the multifractality in case of UrQMD simulated data we have studied the variation of $\frac{d_q}{d_2}$ with order q in figure 3 for $^{16}$O-AgBr, $^{28}$Si-AgBr and $^{32}$S-AgBr interactions. From the slope of the plot of $\frac{d_q}{d_2}$ against q we have calculated the value of **r** which signifies the degree of multifractality. The extracted value **of r** has been presented in table 6 for the UrQMD simulated events. From the table it may be seen that the **r** value characterizing the degree of multifractality calculated for the UrQMD simulated events are significantly lower than the corresponding experimental value for all the three interactions. This observation signifies the presence of stronger multifractality for the experimental data. We have also studied the Ginzburg-Landau phase transition with UrQMD simulated events of $^{16}$O-AgBr, $^{28}$Si-AgBr and $^{32}$S-AgBr interactions. The variation of $\beta_q$ with (q-1) for the UrQMD simulated data have been presented in figure 4, in figure 5 and in figure 6 in case of $^{16}$O-AgBr, $^{28}$Si-AgBr and $^{32}$S-AgBr interactions respectively at 4.5 AGeV/c along with the experimental plot. The variations of $\beta_q$ with (q-1) have been fitted with the function $\beta_q = (q-1)^\nu$ in order to extract the critical exponent $\nu$. In table 7 we have shown the calculated values of the critical exponents for $^{16}$O-AgBr, $^{28}$Si-AgBr and $^{32}$S-AgBr interactions for the experimentally simulated events. From table 7 it may be seen that the critical exponent values for the simulated events are lower than those of the experimental events and there is no evidence of quark-hadron phase transition for the simulated data also. The values of the critical exponent calculated from the multifractal analysis in case of UrQMD simulated data remain almost constant with respect to the mass number of the projectile beam.

We have studied the variation of $D_q$ against $\frac{\ln q}{(q-1)}$ in figure 7 for the UrQMD data sample of $^{16}$O-AgBr, $^{28}$Si-AgBr and $^{32}$S-AgBr interactions in order to calculate the multifractal specific heat. From the slope of the best linear behavior of the plotted points the multifractal specific heat of the shower particles for the UrQMD data sample have been evaluated and presented in table 8 for all the three interactions. From the table it may be seen that the values of the multifractal specific heat for the experimental and simulated data agrees reasonably well with each other. This observation signifies the constancy of multifractal specific heat for the UrQMD simulated events also.

## 6. Conclusions and Outlook

In this paper we have presented an analysis of multifractality and multifractal specific heat in the frame work of Renyi entropy analysis for the produced shower particles in nuclear emulsion detector for $^{16}$O-AgBr, $^{28}$Si-AgBr and $^{32}$S-AgBr interactions at 4.5AGeV/c. Experimental results have been compared with the prediction of Ultra Relativistic Quantum Molecular Dynamics (UrQMD) model. Qualitative information about the multifractal dynamics of particle production process have been extracted and reported in the present analysis. The significant conclusions of this analysis are as follows.

1. Renyi entropy values of all the interactions decrease with the order number q for both experimental and UrQMD simulated data. Renyi entropy values for UrQMD data are less than the corresponding experimental values.

2. Generalized fractal dimension calculated from Renyi entropy for both experimental and UrQMD simulated data decrease with the increase of order q suggesting the presence of multifractality in multi particle production process.

3. The values of Shannon entropy for $^{16}$O-AgBr, $^{28}$Si-AgBr and $^{32}$S-AgBr interactions are little higher than those of Renyi entropies.

4. Degree of multifractality is found to be higher for the experimental data in comparison to the simulated data and it increases with the increase of projectile mass for the experimental data.

5. The experimental values of the critical exponents for $^{16}$O-AgBr and $^{28}$Si-AgBr interactions are lower than the critical value 1.304 required for a quark-hadron phase transition to occur while for $^{32}$S-AgBr interaction the experimentally obtained values of critical exponent is higher than the critical value 1.304 signifying the absence of quark-hadron phase transition. Absence of quark-hadron phase transition is prominent for the simulated events also.
6. The calculated values of critical exponents obtained from our analysis increase with the increase of projectile mass for the experimental data. UrQMD predicted values of the critical exponent $\nu$ remain almost constant with the increase of projectile mass.
7. Multifractal specific heat for the simulated data agrees well with the experimental data. Constancy of multifractal specific heat is reflected from our analysis.

It is true that there are many papers available in the literature where presence of multifractality has been tested experimentally in multiparticle production in high energy nucleus-nucleus interactions by different methods. But the method adopted in this paper to study multifractality seems to be simple and interesting and in this regard our study deserves attention. The observed multifractal behavior of the produced shower particles may be viewed as an experimental fact.

**Acknowledgement**

The authors are grateful to Prof. Pavel Zarubin of the Joint Institute of Nuclear Research (JINR), Dubna, Russia for providing them the required emulsion data. Dr. Bhattacharyya acknowledges Prof. Dipak Ghosh, Department of Physics, Jadavpur University and Prof. Argha Deb Department of Physics, Jadavpur University, for their inspiration in the preparation of this manuscript.

**Table 1**

| Interactions | Average Multiplicity | |
| --- | --- | --- |
| | Experimental | UrQMD |
| $^{16}$O-AgBr (4.5AGeV/c) | 18.05 ± 0.22 | 17.79 ± 0.21 |
| $^{28}$Si-AgBr (4.5AGeV/c) | 23.62 ± 0.21 | 27.55 ± 0.22 |
| $^{32}$S-AgBr (4.5AGeV/c) | 28.04 ± .14 | 30.84 ± 0.17 |

Table 1 represents the average multiplicities of the shower particles for all the interactions in case of the experimental and the UrQMD data.

**Table 2**

| Interactions | Order | Experimental values of Renyi Entropy $H_q$ | UrQMD simulated values of Renyi Entropy $H_q$ |
| --- | --- | --- | --- |
| $^{16}$O-AgBr | 2 | 3.47 ± .04 | 3.05 ± .02 |
| | 3 | 3.41 ± .03 | 2.99 ± .04 |
| | 4 | 3.37 ± .02 | 2.95 ± .02 |
| | 5 | 3.34 ± .02 | 2.92 ± .02 |
| $^{28}$Si-AgBr | 2 | 3.76 ± .05 | 3.29 ± .06 |
| | 3 | 3.69 ± .02 | 3.21 ± .02 |
| | 4 | 3.63 ± .03 | 3.17 ± .02 |
| | 5 | 3.59 ± .03 | 3.13 ± .03 |
| $^{32}$S-AgBr | 2 | 3.91 ± .04 | 3.36 ± .08 |
| | 3 | 3.84 ± .03 | 3.28 ± .03 |
| | 4 | 3.79 ± .03 | 3.24 ± .03 |
| | 5 | 3.74 ± .03 | 3.20 ± .02 |

Table 2 represents the experimental and UrQMD simulated values of Renyi Entropy $H_q$ for $^{16}$O-AgBr, $^{28}$Si-AgBr and $^{32}$S-AgBr interactions at 4.5AGeV/c.

**Table 3**

| Interactions | Order | **Experimental** values of Generalized fractal dimension $D_q$ | **UrQMD** simulated values of Generalized fractal dimension $D_q$ |
|---|---|---|---|
| $^{16}$O-AgBr | 2 | .892±.007 | .826±.002 |
|  | 3 | .877±.006 | .810±.003 |
|  | 4 | .866±.009 | .799±.004 |
|  | 5 | .858±.008 | .791±.005 |
| $^{28}$Si-AgBr | 2 | .897±.004 | .829±.002 |
|  | 3 | .880±.004 | .808±.002 |
|  | 4 | .866±.007 | .798±.004 |
|  | 5 | .857±.007 | .788±.005 |
| $^{32}$S-AgBr | 2 | .924±.004 | .836±.003 |
|  | 3 | .907±.007 | .815±.004 |
|  | 4 | .896±.008 | .805±.005 |
|  | 5 | .884±.008 | .796±.006 |

Table 3 represents the experimental and UrQMD simulated values of Generalized fractal dimension $D_q$ for $^{16}$O-AgBr, $^{28}$Si-AgBr and $^{32}$S-AgBr interactions at 4.5AGeV/c.

**Table 4**

| Interactions | Shannon Entropy S Experimental value | Information dimension Experimental data | Shannon Entropy S UrQMD simulated Value | Information dimension UrQMD data |
|---|---|---|---|---|
| $^{16}O - AgBr$ | $3.59 \pm .12$ | $.922 \pm .001$ | $3.19 \pm .01$ | $.864 \pm .002$ |
| $^{28}Si - AgBr$ | $3.87 \pm .14$ | $.924 \pm .002$ | $3.44 \pm .06$ | $.866 \pm .003$ |
| $^{32}S - AgBr$ | $4.00 \pm .16$ | $.924 \pm .002$ | $3.50 \pm .09$ | $.870 \pm .003$ |

**Table 4 represents the values of Shannon Entropy S and Information dimension for all the three interactions in case of the experimental as well as the UrQMD simulated data.**

**Table 5**

| Interactions | Order | Experimental values of $\frac{d_q}{d_2}$ | Experimental values of $\beta_q$ | UrQMD simulated values of $\frac{d_q}{d_2}$ | UrQMD simulated values of $\beta_q$ |
|---|---|---|---|---|---|
| $^{16}$O-AgBr | 2 | 1.00±.02 | 1.00±.02 | 1.00±.03 | 1.00±.02 |
|  | 3 | 1.14±.03 | 2.28±.03 | 1.09±.02 | 2.18±.02 |
|  | 4 | 1.24±.05 | 3.72±.05 | 1.15±.03 | 3.45±.03 |
|  | 5 | 1.31±.05 | 5.24±.05 | 1.20±.04 | 4.80±.04 |
| $^{28}$Si-AgBr | 2 | 1.00±.02 | 1.00±.02 | 1.00±.02 | 1.00±.02 |
|  | 3 | 1.16±.03 | 2.32±.03 | 1.12±.02 | 2.24±.02 |
|  | 4 | 1.30±.05 | 3.90±.05 | 1.18±.03 | 3.54±.03 |
|  | 5 | 1.39±.06 | 5.56±.05 | 1.24±.04 | 4.96±.04 |
| $^{32}$S-AgBr | 2 | 1.00±.02 | 1.00±.02 | 1.00±.02 | 1.00±.02 |
|  | 3 | 1.22±.04 | 2.44±.03 | 1.13±.03 | 2.26±.02 |
|  | 4 | 1.37±.04 | 4.11±.05 | 1.19±.03 | 3.57±.02 |
|  | 5 | 1.53±.08 | 6.12±.05 | 1.24±.04 | 4.96±.02 |

Table 5 represents the experimental and UrQMD simulated values of $\frac{d_q}{d_2}$ and $\beta_q$ for $^{16}$O-AgBr, $^{28}$Si-AgBr and $^{32}$S-AgBr interactions at 4.5AGeV/c.

**Table 6**

| Interactions | r value characterizing the degree of multifractality obtained from the plot $\frac{d_q}{d_2}$ with order number q | |
|---|---|---|
| | Experimental | UrQMD |
| $^{16}$O-AgBr (4.5AGeV/c) | .206 ± 0.011 | .132 ± 0.006 |
| $^{28}$Si-AgBr (4.5AGeV/c) | .262 ± .011 | .156 ± 0.017 |
| $^{32}$S-AgBr (4.5AGeV/c) | .348 ± .014 | .156 ± 0.019 |

Table 6 represents the **r** value characterizing the degree of multifractality obtained from the plot $\frac{d_q}{d_2}$ with order number q for $^{16}$O-AgBr, $^{28}$Si-AgBr and $^{32}$S-AgBr interactions at 4.5AGeV/c in case of experimental and UrQMD simulated events.

**Table 7**

| Interactions | data | $\nu$ (Calculated From Renyi entropy values) |
|---|---|---|
| $^{16}$O-AgBr interaction at 4.5 AGeV/c | Experimental | 1.197 ± 0.003 |
| | UrQMD | 1.136 ± 0.004 |
| $^{28}$Si-AgBr interaction at 4.5 AGeV/c | Experimental | 1.248 ± 0.010 |
| | UrQMD | 1.152 ± 0.007 |
| $^{32}$S-AgBr interaction at 4.5 AGeV/c | Experimental | 1.327 ± 0.022 |
| | UrQMD | 1.144 ± 0.007 |

Table 7 represents the values of $\nu$ the critical value for the Ginzburg-Landau phase transition for $^{16}$O-AgBr, $^{28}$Si-AgBr and $^{32}$S-AgBr interactions at 4.5 AGeV/c for experimental and UrQMD simulated events calculated from the concept of Renyi entropy.

**Table 8**

| Interactions | Multifractal Specific heat | |
| --- | --- | --- |
| | **Experimental** | **UrQMD** |
| $^{16}$O-AgBr (4.5AGeV/c) | .116 ± 0.004 | .119 ± 0.003 |
| $^{28}$Si-AgBr (4.5AGeV/c) | .138 ± .007 | .138 ± .005 |
| $^{32}$S-AgBr (4.5AGeV/c) | .133 ± .010 | .136 ± .004 |

Table 8 represents the values of multifractal specific heat of the produced shower particles in $^{16}$O-AgBr, $^{28}$Si-AgBr and $^{32}$S-AgBr interactions at 4.5AGeV/c in case of experimental and UrQMD simulated events.

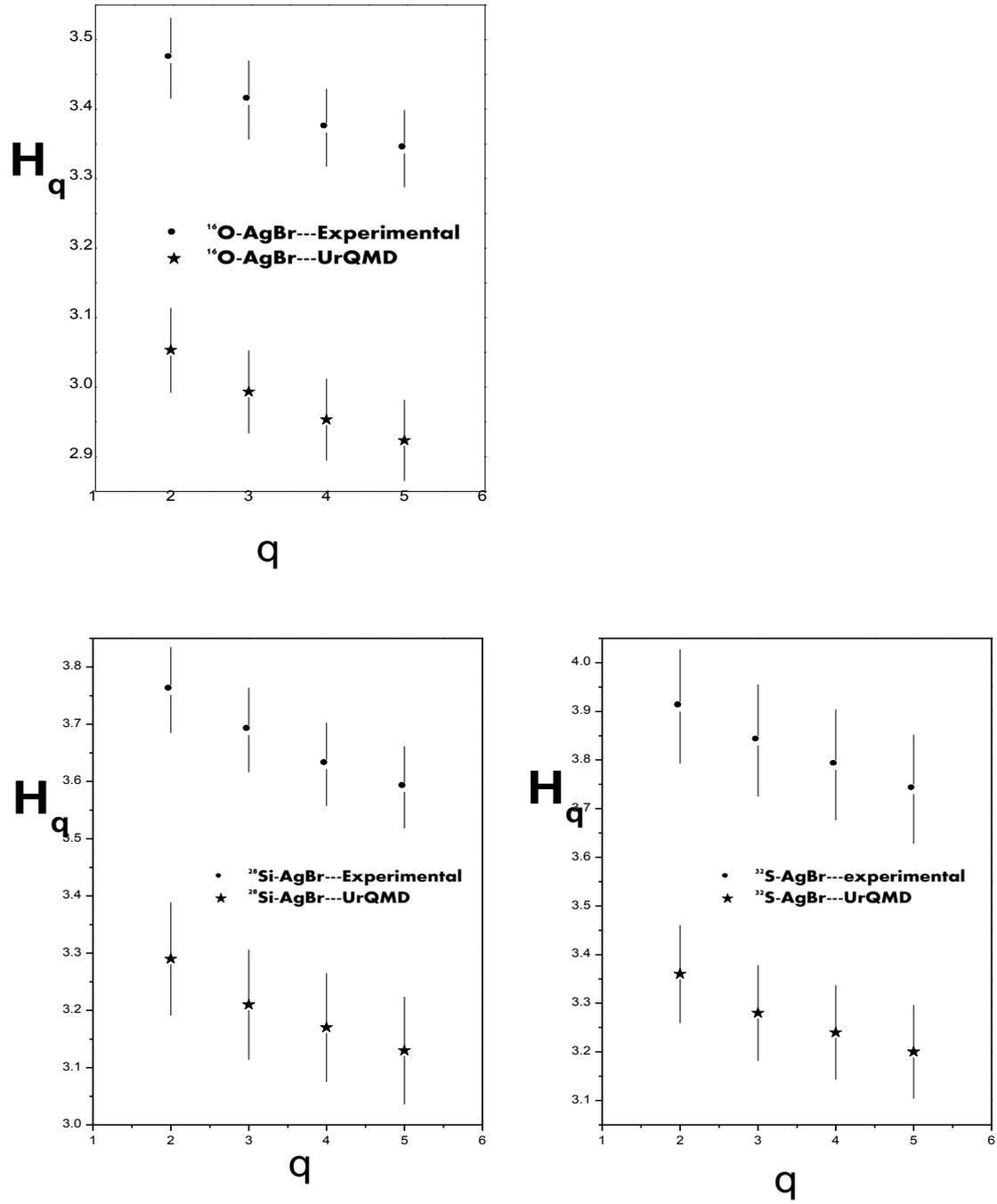

Figure 1 represents the variation of Renyi entropy with order number q for $^{16}$O-AgBr, $^{28}$Si-AgBr and $^{32}$S-AgBr interactions at 4.5AGeV/c in case of experimental and UrQMD simulated events.

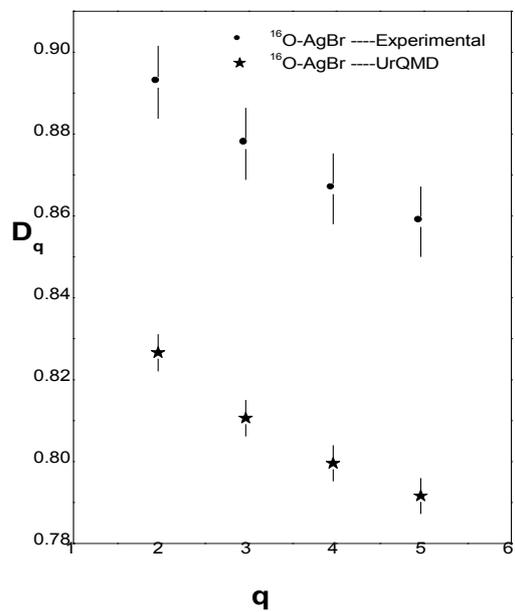

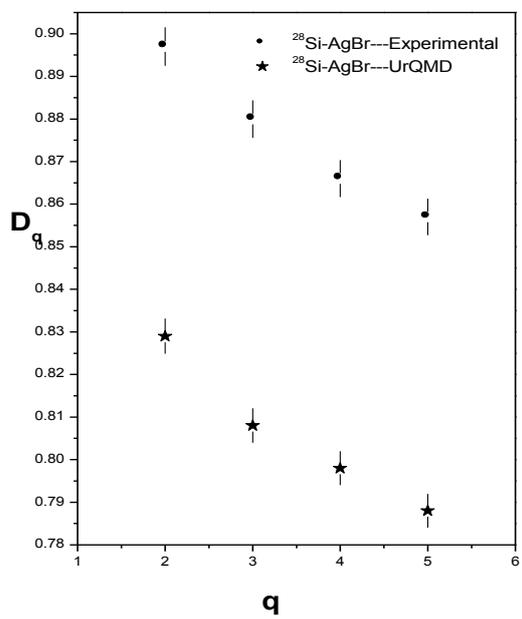
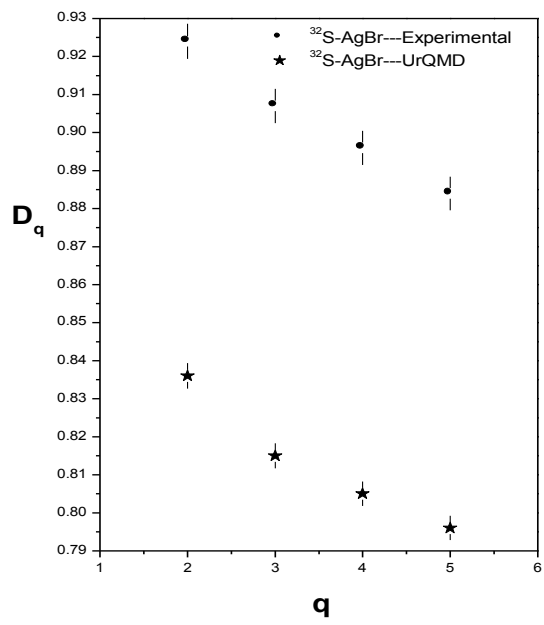

Figure 2 represents the variation of generalized fractal dimension with order number q for $^{16}$O-AgBr, $^{28}$Si-AgBr and $^{32}$S-AgBr interactions at 4.5AGeV/c in case of experimental and UrQMD simulated events.

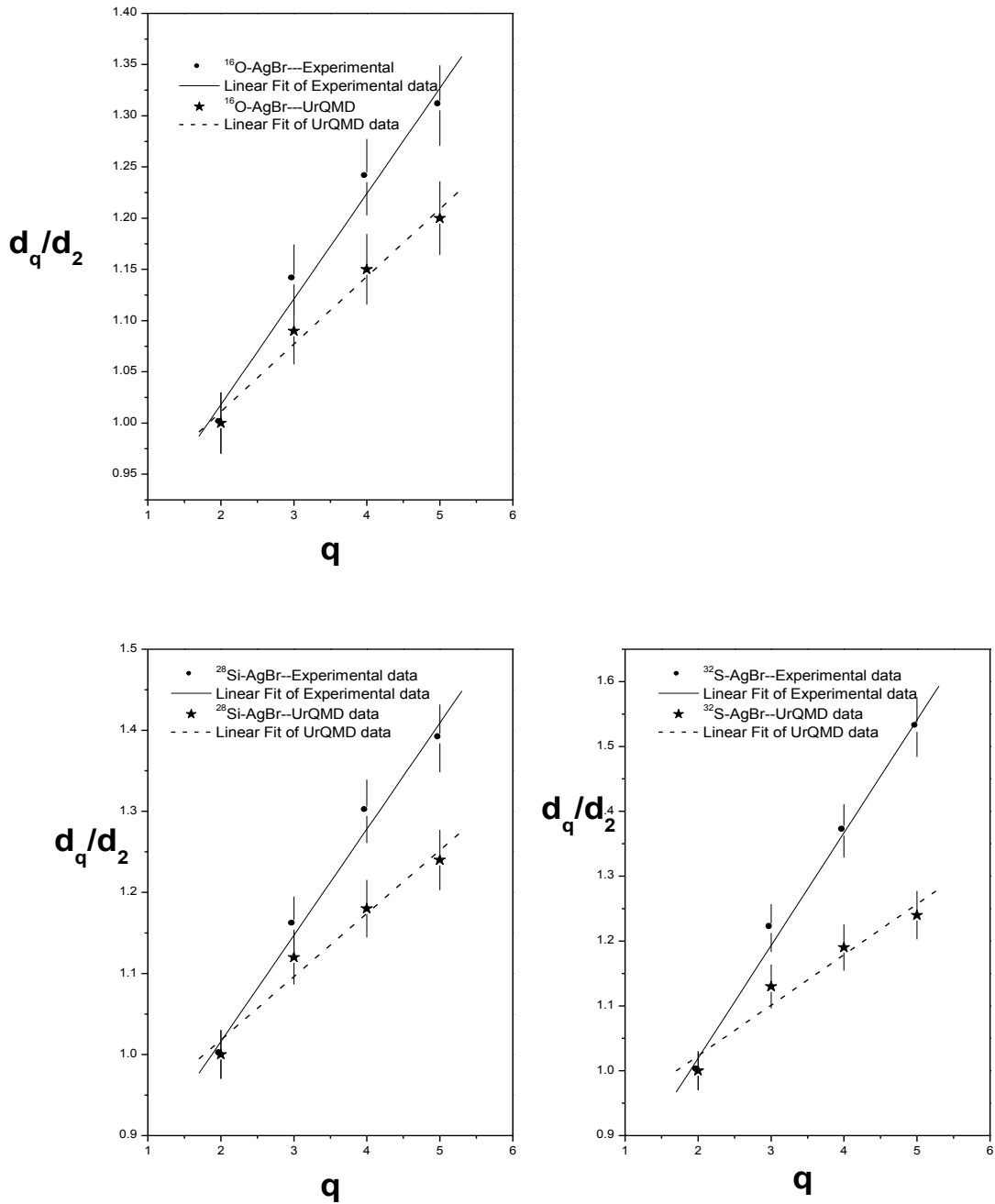

Figure 3 represents the variation of $\frac{d_q}{d_2}$ with order number q for $^{16}$O-AgBr, $^{28}$Si-AgBr and $^{32}$S-AgBr interactions at 4.5AGeV/c in case of experimental and UrQMD simulated events.

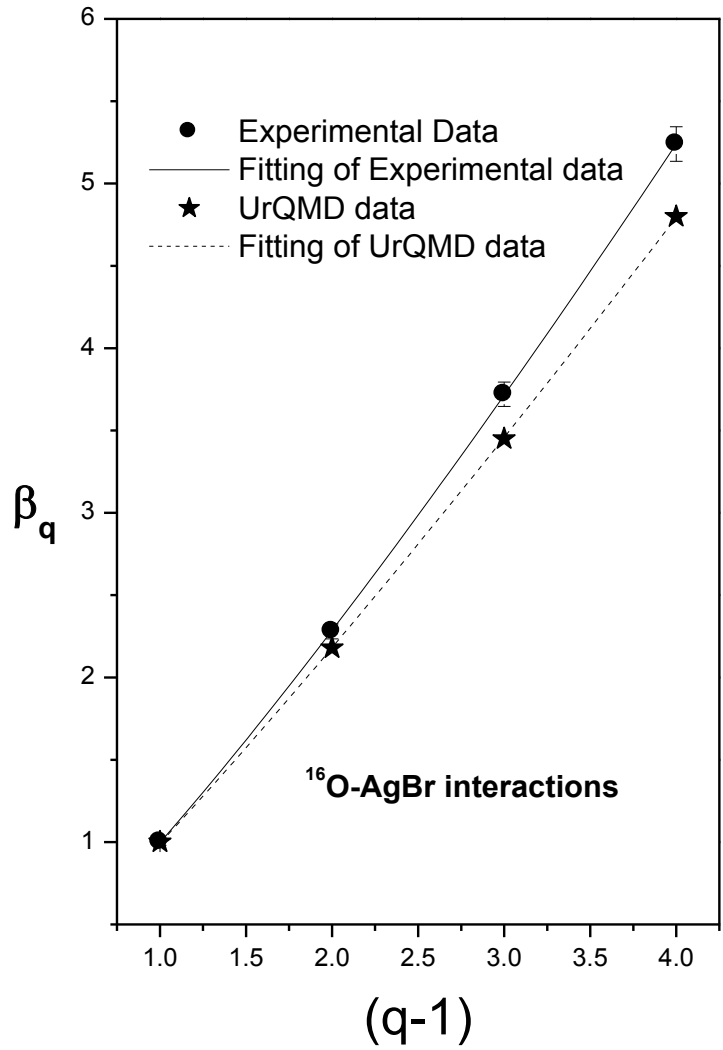

Figure 4 represents the fitting of $\boldsymbol{\beta_q} = (\mathbf{q}-\mathbf{1})^{\boldsymbol{\nu}}$ in case of $^{16}$O-AgBr interactions for the experimental and UrQMD simulated data.

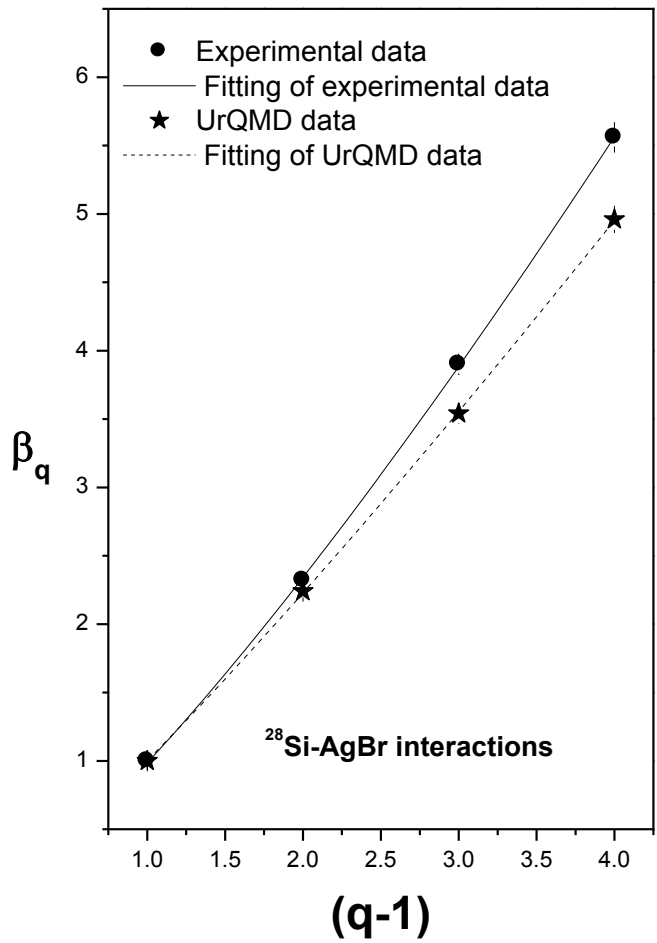

Figure 5 represents the fitting of $\beta_q = (q-1)^\nu$ in case of $^{28}$Si-AgBr interactions for the experimental and UrQMD simulated data.

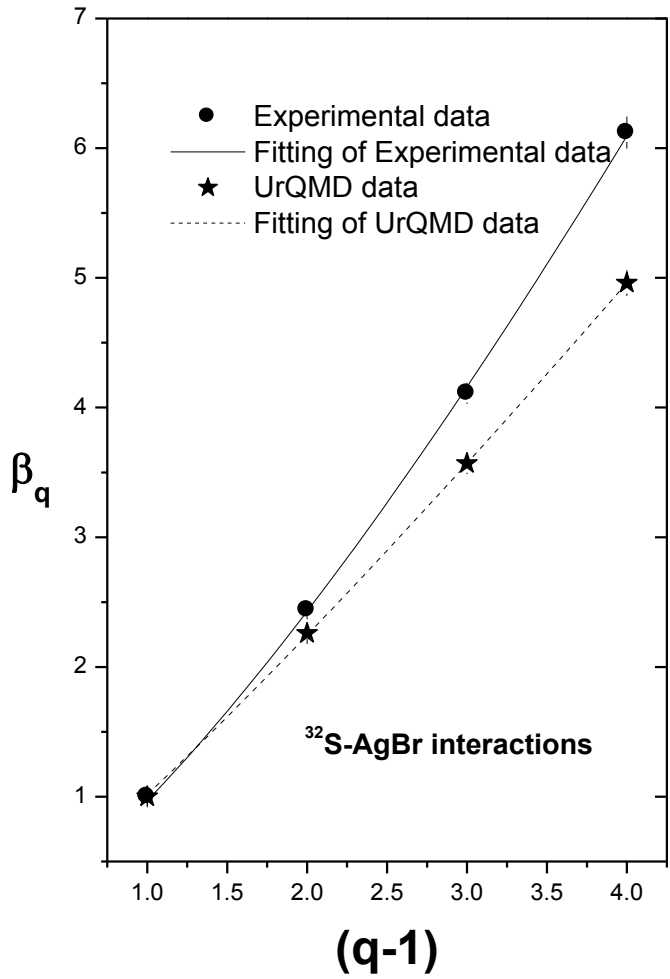

Figure 6 represents the fitting of $\beta_q = (q-1)^\nu$ in case of $^{32}$S-AgBr interactions for the experimental and UrQMD simulated data.

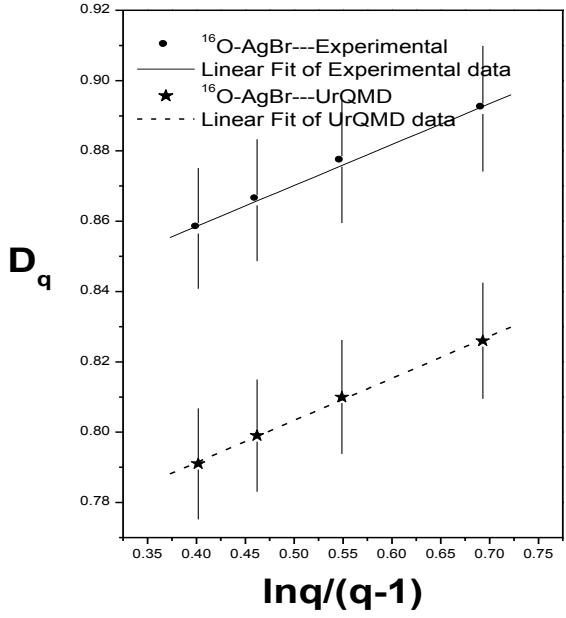

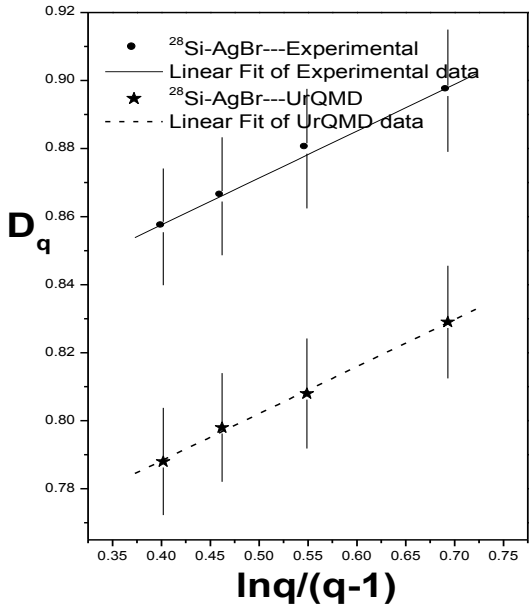
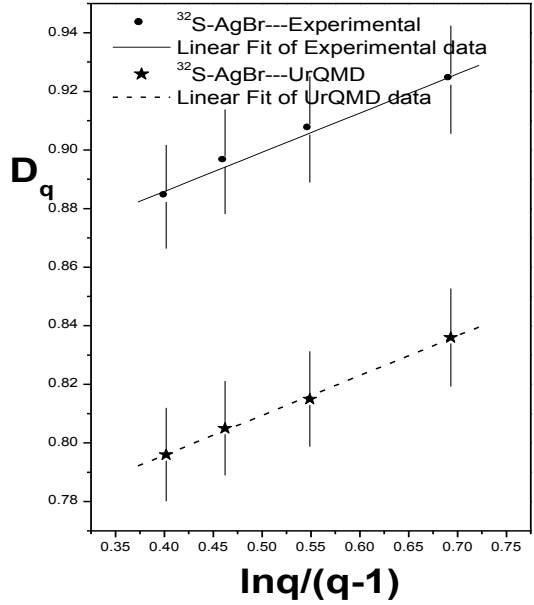

Figure 7 represents the variation of $D_q$ against $\frac{\ln q}{(q-1)}$ for $^{16}$O-AgBr, $^{28}$Si-AgBr and $^{32}$S-AgBr interactions at 4.5AGeV/c in case of experimental and UrQMD simulated events.

There is no conflict of interest in publishing the paper

33